\documentstyle[aps,prl,multicol,epsfig]{revtex}
\begin{document}
\title{ Ground state, quasi-hole, a pair of quasihole wavefunctions and instability in bilayer quantum Hall systems }
\author{ Longhua Jiang and Jinwu Ye }
\address{ Department of Physics, The Pennsylvania State University, University Park, PA, 16802}
\date{\today}
\maketitle
\begin{abstract}
  Bilayer quantum Hall system (BLQH) differ from its single layer counterparts (SLQH)
  by its symmetry breaking ground state
  and associated neutral gapless mode in the pseudo-spin sector.
  Due to the gapless mode, qualitatively good groundstate and low energy
  excited state wavefunctions at any finite distance is still unknown.
  We investigate this important open
  problem by the Composite Boson (CB) theory developed by one of the authors to study BLQH systematically.
  We derive the ground state, quasi-hole and a pair of quasihole
  wavefunctions from the CB theory and its dual action. We find that
  the ground state wavefunction is the product of two parts, one in the charge sector which is
  the well known Halperin's $ (111) $ wavefunction and
  the other in the spin sector which is non-trivial at any finite $ d $ due to the gapless mode.
  So the total  groundstate wavefunction  differs from the
  well known $ (111) $ wavefunction at any finite $ d $.
  In addition to commonly known multiplicative factors, the quasi-hole and a pair of quasi-holes
  wavefunctions also contain
  non-trivial normalization factors multiplying the correct ground state wavefunction.
  We expect that the quasi-hole and pair wave function not only has
  logarithmically divergent energy, well localized charge distribution, but also
  correct interlayer correlations. All the distance dependencies in all the wavefunctions are encoded in
  the spin part of the ground state wavefunction.
  The instability encoded in the spin part of the groundstate wavefunction
  leads to the pseudo-spin density wave formation proposed previously by one of the authors.
  Some subtleties related to the Lowest Landau Level (LLL) projection
  of the CB theory are also noted.
\end{abstract}

\section{Introduction}
   The Wavefunction  approach has been
   very successfully applied to study single layer quantum Hall ( SLQH ) systems at
   Laughlin series $ \nu=\frac{1}{2k+1} $ \cite{laugh} and  Jain's series
   at $ \nu= \frac{ p}{ 2sp \pm 1 }  $ \cite{jain}.
   One of main reasons for the success of
   the wavefunction approach in SLQH is that  there is a gap in the bulk,
   suitable wavefunctions \cite{laugh,jain} can describe both the groundstate and low
   energy excitations quite accurately. Its accuracy can be checked
   easily by exact diagonalization in a finite size system whose size
   need only go beyond a few magnetic length. Spherical geometry can be used to
   get rid of edge state effects quite efficiently. In general, trial wave function approach is a very robust
   approach to study SLQH and multi-components systems as long as
   there is a gap in the bulk. The gap protects the many properties of the system such as charge density
   distributions and energies from being sensitive to some subtle details of wavefunctions.

   However, the situation could be completely different in Spin-polarized Bilayer Quantum Hall
   system ( BLQH ) at total filling factor $ \nu_{T} =1 $. This system has been under enormous experimental and theoretical investigations over the last decade
  \cite{rev}.  When the interlayer separation $ d $ is sufficiently large, the bilayer
  system decouples into two separate compressible $ \nu=1/2 $ layers
  \cite{supp}. However, when $ d $ is smaller than a critical distance $ d_{c1} $, even in the absence
  of interlayer tunneling, the system undergoes a quantum phase transition into
  a novel spontaneous interlayer coherent incompressible phase which
  is an excitonic superfluid state ( ESF ) in the pseudospin channel \cite{gold,hall,counterflow}.
  In \cite{bert}, Halperin proposed the $ (111) $ wavefunction to describe the
  groundstate of the ESF state. Starting from the $ (111) $ wavefunction,
   using various methods, several authors \cite{fer} discovered a
   Neutral Gapless Mode (NGM) with linear dispersion relation $ \omega \sim v k $ and
   that there is a finite temperature Kosterlitz-Thouless (KT)
   phase transition associated with this NGM.
   By treating the two layer indices as two pseudo-spin indices,
   Girvin, Macdonald and collaborators mapped the bilayer system
   into a Easy Plane Quantum Ferromagnet (EPQFM) \cite{rev,moon} ( which is equivalent to
   the ESF )
   and explored many rich and interesting physical phenomena in this system.

   As first pointed out in \cite{moon},
   the $ (111) $ wavefunction may not be qualitatively good at finite $ d $,
   because $ (111) $ is a broken symmetry state in a direction in $ XY $ plane of isotropic ferromagnet at $ d=0 $
   instead of a easy-plane ferromagnet at finite $ d $.
   The NGM is a hallmark of the interlayer coherent Quantum Hall state.
   Its existence is expected to dramatically alter the properties of the wavefunctions of the
   ground state, quasi-hole and quasi-particle.
   In \cite{wave},  G. S. Jeon and one of the authors studied properties of essentially all the known trial wavefunctions of ground state and
   excitations in bilayer quantum Hall systems at the total filling factor $ \nu_{T}=1 $.
   The results indicated that qualitatively good trial wave functions for the ground state and
   the excitations of the interlayer coherent bilayer quantum Hall system at finite
   $ d $ are still not available and searching for them remains an important open problem.
   Specifically, they investigated the properties of
   the quasi-hole wave function, meron wavefunction and a pair of
   meron wavefunction  built on $ (111) $ state which have superscripts "prime" in this paper:
\begin{equation}
  \Psi^{\prime}_{qh}= ( \prod^{N_{1}}_{i}z_{i} ) \Psi_{111},~~~
  \Psi^{\prime}_{meron}= ( \prod^{N_{1}}_{i} \frac{ z_{i}}{|z_{i} |
  } ) \Psi_{111},~~~ \Psi^{\prime}_{pair} = \prod^{N_{1}}_{i} ( z_{i}- z_{0})
  \prod^{N_{2}}_{i}( w_{i}- w_{0} ) \Psi_{111}
\label{prime}
\end{equation}
  where $ \Psi_{111} $ is the Halperin's $ (111) $ wavefunction:
\begin{equation}
  \Psi_{111} (z,w)= \prod^{N_{1}}_{ i< j } ( z_{i}-z_{j} )   \prod^{N_{2}}_{ i< j } (
  w_{i}-w_{j} ) \prod^{N_{1}}_{i=1} \prod^{N_{2}}_{j=1} (z_{i}-w_{j} )
  \exp(-\frac{1}{4l^{2}_{0}}\sum_{i}|z_{i}|^{2}+|w_{i}|^{2})
\label{111}
\end{equation}
    where $ N_{1}=N_{2} = N $ in the balanced case and
    $ z $ and $ w $ are the coordinates in layer 1 and layer 2 respectively.
    In the following, we suppress the exponential factor.

  These quasi-hole and meron wavefunctions differ only by "normalization
  factors". As shown in \cite{wave}, the normalization factor $ |z_{i} | $ is accurate only at long distance
  limit $ | z_{i} | \rightarrow \infty $ limit. Near the origin, the "meron" and the "quasi-hole ' have similar
  behaviors. Normalization factors have been shown not to be important in single
  layer Quantum Hall ( SLQH ) systems.
  However, as shown in \cite{wave}, they make a dramatic difference in BLQH.
  Although the smallest meron has
  a localized charge $ 1/2 $ and logarithmically divergent energy,
  the charge of the quasi-hole excitation extends
  over the whole system and its energy also diverges linearly as the area of the system size.
  This indicates that the
  quasi-hole wavefunction is not a good trial wavefunction for any low energy excitations.
  The meron wavefunction is  not a good trial wavefunction either,
  because it ignores the strong interlayer correlations \cite{wave}.
  It was found the energy of the possible wavefunction of a pair of merons
  in Eqn.\ref{prime} increases quadratically $ \sim | z_{0} -w_{0} |^{2} $
  instead of logarithmically as the separation of the pair increases.
  All the results achieved in \cite{wave} indicate that
  qualitatively good trial wave functions in the interlayer coherent
  bilayer quantum Hall system at finite $ d $, both for ground state and excitations, are still
  unknown and searching for them remains an important open problem.
  so the wavefunction approach to BLQH is much more difficult and far less powerful in BLQH than in SLQH.
  Fortunately, effective theory approaches such as EPQFM approach \cite{moon,rev}  and  Composite Boson
  theory approach \cite{fer,moon,cbprl,cbprlim,cbtwo} circumvent this difficulty associated with
  the unknown wavefunction at any finite $ d $ and are very effective
  to bring out most of the interesting phenomena in the pseudo-spin sector in this system.
  In fact, all these effective theories start from the insights gained from Halperin's $ (111) $
  wavefunction which is exact at $ d=0 $.

  In a series of papers \cite{cbprl,cbprlim,cbtwo}, one of the
  authors developed a systematic composite boson approach to study
  balanced and im-balanced Bi-Layer Quantum Hall systems in rather
  details. The theory  puts spin and charge degree freedoms in the same footing,
  explicitly bring out the spin-charge connection and classify all the possible excitations in
  a systematic way. Then he pushed the theory further to understand
  novel phases and quantum phase transitions as the distance
  between the two layers is changed. He found that starting from the well
  studied excitonic superfluid (ESF) state, as distance increases,
  the instability driven by magneto-roton minimum collapsing at a finite wavevector in the pseudo-spin channel
  leads to the formation of a pseudo-spin density wave (PSDW) at
  some intermediate distances. He constructed a quantum
  Ginsburg-Landau theory  to study the transition from the excitonic superfluid (ESF) to the
  PSDW and analyze in detail the properties of the PSDW.
  He showed that a square lattice is the favorite lattice and
  the correlated hopping of vacancies in the active and passive layers in the PSDW
  state leads to very large and temperature dependent drag observed in the
  experimental. In the presence of disorders, the properties of the PSDW are
  consistent with all the experimental observations \cite{hall,drag} in the intermediate distances.
  Further experimental implications  of the PSDW are given.
  Then he extended the Composite Boson theory to study slightly im-balanced
  BLQH. In the global $ U(1) $ symmetry breaking
  excitonic superfluid side, as the imbalance increases,
  the system supports continuously changing fractional charges.
  In the translational symmetry breaking PSDW side, there are two quantum
  phase transitions from the commensurate PSDW to an
  in-commensurate PSDW and then to the excitonic superfluid state.
  These results explained the experimental observations in
  \cite{imbexp} very nicely.
  The author found that the theory can be easily extended to study
  some additional interesting phenomena in trilayer quantum Hall
  systems \cite{tricb}. It was concluded in \cite{cbtwo} that
  field theory approaches are much more powerful in BLQH than in SLQH.

  Obviously, the CB theory circumvent this difficulty associated with
  the unknown wavefunction at any finite $ d $ and is used to achieve
  all these interesting and important results  at two different distance regimes without knowing the precise wavefunctions
  for the ground state and excitations. It would be interesting to
  use the CB theory to address the important and outstanding problem
  avoided in \cite{wave} and in all the other pervious work that
  finding the good ground state and low energy excited wavefunction for BLQH at
  any finite $ d $. In SLQH, the CB theory developed in \cite{cb}
  was used to re-derive the already well known Laughlin's wavefunctions
  for ground state and quasihole at $ \nu=\frac{1}{2k+1} $. As said previously, the gap in the bulk protects
  the properties of the system such as charge density
  distributions and energies from being sensitive to some subtle details of
  wavefunctions.  Here we are facing a more
  difficult and interesting task: to derive these unknown
  wavefunctions at finite $ d $.

  The rest of the paper is organized
  as the following. In Sec. II, in order to be self-contained, we
  review briefly the CB approach and its dual action developed in \cite{cbtwo} which are needed to
  derive the wavefunctions in the following sections. In Sec. III, using the formalism
  presented in Sec. II, we derive the ground state wavefunction
  which is different from the $ (111) $ wavefunction at any finite $
  d $. In Sec.IV, using the dual action presented in Sec. II, we
  derive the quasi-hole wavefunction and compare it with the
  "quasi-hole" and "meron" wavefunction built on $ (111) $
  wavefunction listed in Eqn.\ref{prime}. In Sec.V, we derive a pair of
  meron wavefunction with charge $ 1 $ and compare it with the "
  pair meron " wavefunction built on $ (111) $ listed in
  Eqn. \ref{prime}. In Sec.VI, we look at the instability in the ground
  state wavefunction as distance approaches $ d_{c1} $. Finally, we
  reach conclusions in Sec. VII. Some caveats related to the Lowest Landau Level (LLL) projection
  of the wavefunctions are also pointed out.  We note that there is also an
  alternative approach in \cite{mixed}.

\section{ Composite boson approach and its dual action in BLQH}

 In this section, we briefly review the formalism developed in
 \cite{cbtwo} which is needed to derive the wavefunctions in the
 following sections. Consider a bi-layer system with $ N_{1} $ ( $ N_{2} $ ) electrons in
 left ( right ) layer and
  with interlayer distance $ d $ in the presence of magnetic field $ \vec{B} = \nabla \times \vec{A} $ ( Fig.1):
\begin{eqnarray}
   H & = & H_{0} + H_{int}    \nonumber  \\
   H_{0} & = &  \int d^{2} x c^{\dagger}_{\alpha}(\vec{x})
   \frac{ (-i \hbar \vec{\nabla} + \frac{e}{c} \vec{A}(\vec{x}) )^{2} }
      {2 m }  c_{\alpha}(\vec{x})                \nonumber   \\
    H_{int} & = &  \frac{1}{2} \int d^{2} x d^{2} x^{\prime} \delta \rho_{\alpha} (\vec{x} )
               V_{\alpha \beta} (\vec{x}-\vec{x}^{\prime} )  \delta \rho_{\beta } ( \vec{x^{\prime}} )
\label{first}
\end{eqnarray}
  where electrons have {\em bare} mass $ m $ and carry charge $ - e $, $ c_{\alpha}, \alpha=1,2 $ are
  electron  operators in top and bottom layers, $ \delta \rho_{\alpha}(\vec{x}) = c^{\dagger}_{\alpha} (\vec{x})
  c_{\alpha} (\vec{x} ) -n_{\alpha}, \alpha=1,2 $ are normal ordered electron densities on each layer.
  The intralayer interactions
  are $ V_{11}=V_{22}= e^{2}/\epsilon r $, while interlayer interaction is $ V_{12}=V_{21}= e^{2}/ \epsilon
  \sqrt{ r^{2}+ d^{2} } $ where $ \epsilon $ is the dielectric constant.

    Performing a singular gauge transformation \cite{cbprl,cbtwo}:
\begin{equation}
  \phi_{a}(\vec{x}) = e ^{ i \int d^{2} x^{\prime} \phi (\vec{x}-\vec{x}^{\prime} )
  \rho ( \vec{x}^{\prime} ) } c_{a}( \vec{x})
\label{singb}
\end{equation}
     where $ \phi( \vec{x}-\vec{x}^{\prime} )= arg(\vec{x}-\vec{x}^{\prime} ) $ is the angle between
     the vector $ \vec{x}-\vec{x}^{\prime} $ and the horizontal axis.
     $ \rho ( \vec{x} ) = c^{\dagger}_{1}( \vec{x} ) c_{1}( \vec{x} ) +
     c^{\dagger}_{2}( \vec{x} ) c_{2}( \vec{x} )  $ is the total density of the bi-layer system.
     Note that this transformation treats both $ c_{1} $ and $ c_{2} $ on the same footing.
     This is reasonable only when the distance between the two layers is sufficiently small.
     It can be shown that $ \phi_{a}(\vec{x}) $ satisfies all the boson commutation relations.
     We can transform the Hamiltonian Eqn.\ref{first} into the Lagrangian in Coulomb gauge:
\begin{eqnarray}
   {\cal L} & = & \phi^{\dagger}_{a}( \partial_{\tau}- i a_{0} ) \phi_{a}
    + \phi^{\dagger}_{a}(\vec{x}) \frac{ (-i \hbar \vec{\nabla} + \frac{e}{c} \vec{A}(\vec{x})
             - \hbar \vec{ a }(\vec{x}) )^{2} }{2 m }  \phi_{a}(\vec{x})
                      \nonumber  \\
     & + & \frac{1}{2} \int d^{2} x^{\prime} \delta \rho (\vec{x} )
               V_{+} (\vec{x}-\vec{x}^{\prime} )  \delta \rho ( \vec{x^{\prime}} )
                           \nonumber  \\
    & + & \frac{1}{2} \int d^{2} x^{\prime} \delta \rho_{-} (\vec{x} )
               V_{-} (\vec{x}-\vec{x}^{\prime} )  \delta \rho_{-} ( \vec{x^{\prime}} )
    -\frac{ i }{ 2 \pi} a_{0} ( \nabla \times \vec{a} )
\label{boson}
\end{eqnarray}
   where $ V_{\pm}= \frac{ V_{11} \pm V_{12} }{2} $ and $ V_{11}= V_{22}=\frac{2 \pi e^{2}}{\epsilon q },
    V_{12}= \frac{2 \pi e^{2}}{\epsilon q } e^{-qd} $. The
    Chern-Simon gauge field is $ \vec{a}= \int d^{2}
    \vec{r}^{\prime} \nabla \phi( \vec{x}-\vec{x}^{\prime} ) \rho(
    \vec{x}^{\prime} )= \int d^{2} \vec{r}^{\prime}
    \frac{ \hat{z} \times ( \vec{x}-\vec{x}^{\prime} )}{ |\vec{x}-\vec{x}^{\prime} |^{2} } \rho(
    \vec{x}^{\prime} ) $.

   In Coulomb gauge, integrating out $ a_{0} $ leads to the constraint: $ \nabla \times \vec{a}
   = 2 \pi  \phi^{\dagger}_{a} \phi_{a}  $. Note that if setting $
   V_{-} =0 $, then the above equation is identical to a single
   layer with spin in the absence of Zeeman term, so the Lagrangian
   has a $ SU(2) $ pseudo-spin symmetry. The $ V_{-} $ term breaks
   the $ SU(2) $ symmetry into $ U(1) $ symmetry. In the BLQH at finite $ d $, $
   V_{-} > 0 $, so the system is in the Easy-plane limit.

    We can write the two bosons in terms of magnitude and phase
\begin{equation}
  \phi_{a}= \sqrt{ \bar{\rho}_{a} + \delta \rho_{a} } e^{i \theta_{a} }
\label{decom}
\end{equation}

   The boson commutation relations imply that
   $ [ \delta \rho_{a} ( \vec{x} ), \theta_{b}( \vec{x} ) ] =
     i \hbar \delta_{ab} \delta( \vec{x}-\vec{x}^{\prime} ) $.
   After absorbing the external gauge potential $ \vec{A} $ into $
   \vec{a} $, we get the Lagrangian in the Coulomb gauge:
\begin{eqnarray}
  {\cal L}  &  =  & i \delta \rho^{+} ( \frac{1}{2} \partial_{\tau} \theta^{+}-  a_{0} ) +
          \frac{ \bar{\rho} }{2m} [ \frac{1}{2} \nabla \theta_{+} + \frac{1}{2} (\nu_{1}-\nu_{2} ) \nabla \theta_{-}
          - \vec{a} ]^{2}
     +  \frac{1}{2} \delta \rho^{+} V_{+} (\vec{q} )  \delta \rho^{+}
         - \frac{ i }{ 2 \pi} a_{0} ( \nabla \times \vec{a} )
                       \nonumber   \\
    & + & \frac{i}{2} \delta \rho^{-}  \partial_{\tau} \theta^{-} +
          \frac{ \bar{\rho} f }{2m}  ( \frac{1}{2} \nabla \theta_{-} )^{2}
          + \frac{1}{2} \delta \rho^{-} V_{-} (\vec{q} )  \delta \rho^{-}
          - h_{z} \delta \rho^{-}
\label{main}
\end{eqnarray}
    where $ f= 4 \nu_{1} \nu_{2} $ which is equal to 1 at the balanced case and
    $ h_{z}= V_{-} \bar{\rho}_{-} = V_{-} ( \bar{\rho}_{1}-  \bar{\rho}_{2} ) $ plays a role like a Zeeman field.

   Performing the duality transformation on Eqn.\ref{main} leads to the dual action in terms of
   the vortex degree of freedoms $ J^{v \pm}_{\mu} = \frac{1}{ 2 \pi} \epsilon_{\mu \nu \lambda}
   \partial_{\mu} \partial_{\nu} \theta_{\pm} = J^{v1}_{\mu} \pm J^{v2}_{\mu} $ and the corresponding
   dual gauge fields $ b^{\pm}_{\mu} $:
\begin{eqnarray}
   {\cal L}_{d} & = & - i \pi b^{+}_{\mu} \epsilon_{\mu \nu \lambda} \partial_{\nu} b^{+}_{\lambda}
   - i A^{+}_{ s \mu} \epsilon_{\mu \nu \lambda} \partial_{\nu} b^{+}_{\lambda}
    + i \pi b^{+}_{\mu} J^{v+}_{\mu}
    +  \frac{ m }{ 2 \bar{\rho} f } ( \partial_{\alpha} b^{+}_{0}
      - \partial_{0} b^{+}_{\alpha} )^{2} + \frac{1}{2} ( \nabla \times \vec{b}^{+} ) V_{+} ( \vec{q})
      ( \nabla \times \vec{b}^{+} )
                            \nonumber   \\
   & - &  i A^{-}_{ s \mu} \epsilon_{\mu \nu \lambda} \partial_{\nu} b^{-}_{\lambda}
    + i \pi b^{-}_{\mu} J^{v-}_{\mu}
     - h_{z} ( \nabla \times \vec{b}^{-} )
    +   \frac{ m }{ 2 \bar{\rho} f } ( \partial_{\alpha} b^{-}_{0}
      - \partial_{0} b^{-}_{\alpha} )^{2} + \frac{1}{2} ( \nabla \times \vec{b}^{-} ) V_{-} ( \vec{q})
      ( \nabla \times \vec{b}^{-} )
                      \nonumber   \\
   & - & \frac{m}{ \bar{\rho} f } ( \nu_{1}-\nu_{2} )
   ( \partial_{\beta} b^{-}_{0} - \partial_{0} b^{-}_{\beta} )
      ( \partial_{\beta} b^{+}_{0} - \partial_{0} b^{+}_{\beta} )
\label{dual}
\end{eqnarray}
     where $ A^{\pm}_{s \mu}= A^{1}_{ s \mu} \pm  A^{2}_{ s \mu} $ are the two source fields.
     It is useful to stress that the dual CS term only appears in the
     charge sector.

    For simplicity, we only consider the balanced case.
    Putting  $ \nu_{1}= \nu_{2}= 1/2 $ and $ h_{z}= 0 $ into Eqn.\ref{main}, we get the
    Lagrangian in the balanced case where the symmetry is enlarged to $ U(1)_{L} \times U(1)_{G} \times Z_{2} $.

\section{ Ground state wavefunction }

  In this section, we derive the ground state wavefunction from the
  formalism reviewed in the last section.
  From Eqn.\ref{main}, we can find the corresponding Hamiltonian in the charge sector:
 \begin{equation}
    {\cal H}_{c} = \frac{1}{2} \sum_{q} [ 4 \Pi_{+} (- \vec{q} ) ( V_{+}( q) + \frac{ 4 \pi^{2} \bar{\rho} }{m}
     \frac{1}{ q^{2} } ) \Pi_{+} ( \vec{q} ) + \frac{ \bar{\rho} q^{2} }{ 4 m } \theta_{+} ( - \vec{q} )
     \theta_{+} (\vec{q} ) ]
\label{ch}
\end{equation}
    where $ \Pi_{+}( \vec{q} ) =\delta \rho^{+}/2 $ and
    $ [ \theta_{+} ( \vec{q} ), \Pi_{+} ( \vec{q}^{\prime} ) ] = i \hbar \delta( \vec{q} + \vec{q}^{\prime} ) $.

    Representing $ \theta_{+}  (\vec{q} ) $ by $ -i \frac{ \partial}{ \partial \Pi_{+}( - \vec{q} ) } $,
    in the long wavelength limit, neglecting the  Coulomb
    interaction $ V_{+}(q) \sim 1/q $ which is less singular than $
    1/q^{2} $, we find the ground state wavefunction in the charge sector:
\begin{equation}
    \Psi^{b}_{c0} [ \Pi_{+} ( \vec{q} ) ]  = \exp [ -  \frac{1}{2} \sum_{q} \frac{ 2 \pi }{ q^{2} }
     \delta \rho_{+} (- \vec{q} ) \delta \rho_{+} ( \vec{q} ) ]
\end{equation}

     In the following, we use $ \vec{x} $ to stand for the complex coordinate $ x+iy $.
     Using $ \delta \rho_{+} (x) = \sum \delta( x- z_{i} ) + \sum \delta( x- w_{i} ) - \bar{\rho} $ and
     transforming to the position space, it is simply the modulus of the $ ( 111) $
     wavefunction in the  balanced case $  \Psi^{b}_{c0}=|\Psi_{111} | $. This is the wavefunction in the bosonic picture.
     In order to get the wavefunction in the original fermionic picture, we need to perform the inverse
     of the SGT in Eqn.\ref{singb} on the bosonic wavefunction.
     In the first quantization form, the inverse is:
\begin{equation}
  U_{0} = \prod^{N_{1}}_{ i< j } \frac{ ( z_{i}-z_{j} )}{ |z_{i}-z_{j} | }
      \prod^{N_{2}}_{ i< j } \frac{(w_{i}-w_{j} )}{ |w_{i}-w_{j} | }
      \prod^{N_{1}}_{i=1} \prod^{N_{2}}_{j=1} \frac{ (z_{i}-w_{j})}{ | z_{i}-w_{j} | }
\label{singg}
\end{equation}
     Performing the inverse transformation on the modulus leads to the $ ( 111) $
     wavefunction in the fermionic coordinates:
\begin{equation}
  \Psi_{c0}= U_{0} \Psi^{b}_{c0}= \Psi_{111} (z,w)
\end{equation}

    In contrast to the SLQH, there is also an additional pseudo-spin
    sector in the BLQH which contains the most interesting physics.
    From Eqn.\ref{main}, we can find the corresponding Hamiltonian in the spin sector is:
 \begin{equation}
    {\cal H}_{s} = \frac{1}{2} \sum_{q} [ 4 V_{-} \Pi_{-} (- \vec{q} ) \Pi_{-} ( \vec{q} ) +
     \frac{ \rho_{E}  q^{2} }{ 4 } \theta_{-} ( - \vec{q} ) \theta_{-} (\vec{q} ) ]
\label{spinh}
\end{equation}
    where $ \Pi_{-} ( \vec{q} ) = \delta \rho^{-}/2 $ and
    $ [ \theta_{-} ( \vec{q} ), \Pi_{-} ( \vec{q}^{\prime} ) ] = i \hbar \delta( \vec{q} + \vec{q}^{\prime} )
    $; $ \rho_{E}= \bar{\rho}/m $ is the spin stiffness \cite{compare}. At small $ q $, $
    V_{-}(q)= a-bq+ c q^{2} $ \cite{compare}  where $ a \sim d^{2}, b \sim d^{2} $
    and $ c $ remains a constant at small distances \cite{moon,cbtwo}.
    It is important to stress that this form of $ V_{-} (q) $ has
    the shape displayed in the Fig.1, it not only has a phonon part
    near $ q=0 $, also has a roton part near $ q=q_{0} \sim  1/l_{B}
    $ where $ l_{B} $ is the magnetic length.

    Representing $ \theta_{-}  (\vec{q} ) $ by $ -i \frac{ \partial}{ \partial \Pi_{-}( - \vec{q} ) } $,
    we find the ground state wavefunction in the spin sector:
\begin{equation}
    \Psi_{s0} [ \Pi_{-} ( \vec{q} ) ]  = \exp [ -  \frac{1}{2} \sum_{q} \frac{ \sqrt{ V_{-}(q)/\rho_{E} } }{ q }
    \delta \rho_{-} (- \vec{q} ) \delta \rho_{-} ( \vec{q} ) ]
\label{spinw}
\end{equation}
     It is easy to see that the above equation make senses only when $ V_{-}(q) $ is
     positive for all $ q $.

     At $ d=0 $, $ a=b =0 $, $ V_{-}(q) = c q^{2} $, then Eqn.\ref{spinw}
     becomes:
\begin{equation}
    \Psi_{s0} [ \Pi_{-} ( \vec{q} ); d=0 ]  = \exp [ -  \frac{1}{2} \sum_{q}  \sqrt{ c /\rho_{E} }
    \delta \rho_{-} (- \vec{q} ) \delta \rho_{-} ( \vec{q} )
    ]=const.
\label{spinw0}
\end{equation}

    At any finite distance, as long as  $ d < d_{c1} \sim l_{B} $ in the Fig.1, so the roton has a large gap,
    we can neglect the contributions from the roton part in Fig.1 and only focus on the phonon contributions.
    In the long wavelength limit $ q \ll q_{0} \sim 1/l_{B} $, $  V_{-}(q) \rightarrow a \sim
    d^{2} $.
    Using $ \delta \rho_{-} ( x ) = \sum \delta( x- z_{i} ) - \sum
    \delta( x- w_{i} ) $ and transforming to the coordinate space, it is
\begin{equation}
    \Psi_{s0} ( z, w )  = \exp[ - 2 \sqrt{ 1 /\rho_{E} } (  \sum_{i < j} \frac{ d }{ | z_{i} - z_{j} | }
    - \sum_{i , j} \frac{ d }{ | z_{i} - w_{j} | } + \sum_{i < j} \frac{ d }{ | w_{i} - w_{j} | } )  ]
\label{spinw1}
\end{equation}

     Obviously, the above equation only holds  at small distance $ d < d_{c1} $ and
     in the long distance limit $ | z_{i} - w_{j} | \gg l_{B} $ \cite{frad}.

     The total wavefunction is:
\begin{equation}
    \Psi_{0} ( z, w ) = \Psi_{111} (z,w) \Psi_{s0} ( z, w )
\label{totalg}
\end{equation}

     It is easy to see that the total wavefunction coincides with the $ (111) $ wavefunction only in $ d \rightarrow 0 $ limit.
     At any finite $ d $, it has an extra factor from the gapless spin sector $ \Psi_{s0} ( z, w ) $.
     Note that this extra spin factor Eqn.\ref{spinw1} is not in the
     LLL, this should not be too worrisome, because similar to the
     meron wavefunction listed in Eqn.\ref{prime}, modulus of the
     coordinates could appear in the long distance limit where
     Eqn.\ref{spinw1} hold.

\section{ Quasi-hole wave functions}

  By inserting one static vortex at the origin in layer 1 by setting $ J^{+v}_{0} = J^{-v}_{0} = \delta(x) $
  or layer 2 by setting $ J^{+v}_{0} = - J^{-v}_{0} = \delta(x) $, from the dual action Eqn.\ref{dual},
  we will first try to derive the quasi-hole wavefunction and   compare this quasi-hole wavefunction with
  the known "quasi-hole " wavefunction and "meron" wavefunction written down in \cite{wave}.

    In order to derive the quasi-hole wavefunction, we have to resort to the dual action Eqn.\ref{dual} in the
    balanced case where the last term vanishes. Setting the two sources $ A^{\pm}_{s \mu} =0 $,
    in the Coulomb gauge $ \nabla \cdot b^{\pm}_{\alpha} =0 $,
    Eqn.\ref{dual} becomes:
\begin{eqnarray}
   {\cal L}_{d} & = & - i 2 \pi b^{+}_{0} \epsilon_{\alpha \beta } \partial_{\alpha} b^{+}_{\beta}
    + i \pi b^{+}_{0} J^{v+}_{0} + i \pi b^{+}_{\alpha} J^{v+}_{\alpha}
    +  \frac{ m }{ 2 \bar{\rho}  } ( \partial_{\alpha} b^{+}_{0}
    )^{2} + \frac{ m }{ 2 \bar{\rho}  } ( \partial_{0} b^{+}_{\alpha}
    )^{2}  + \frac{1}{2} ( \nabla \times \vec{b}^{+} ) V_{+} ( \vec{q})
      ( \nabla \times \vec{b}^{+} )
                            \nonumber   \\
   & + & i \pi b^{-}_{0} J^{v-}_{0} + i \pi b^{-}_{\alpha}
   J^{v-}_{\alpha} + \frac{ m }{ 2 \bar{\rho}  } ( \partial_{\alpha} b^{-}_{0}
    )^{2} + \frac{ m }{ 2 \bar{\rho}  } ( \partial_{0} b^{-}_{\alpha}
    )^{2} + \frac{1}{2} ( \nabla \times \vec{b}^{-} ) V_{-} ( \vec{q})( \nabla \times \vec{b}^{-} )
\label{dualc}
\end{eqnarray}
    Note the absence of CS term in the spin sector.

  We only consider static vortices, so $ J^{v+}_{\alpha}=  J^{v-}_{\alpha}=0 $.
  Integrating out $b^{+}_{0}$ and $b^{-}_{0}$ and transforming into the coordinate
  space lead to:
\begin{eqnarray}
{\cal L}_{d}&=&-{1\over 4 \pi}\frac{ \bar{\rho}  }{2m}(\pi J^{v+}_{0}-2\pi \epsilon_{\alpha,\beta}
\partial_{\alpha}b^{+}_{\beta})ln|x-y|(\pi J^{v+}_{0}-2\pi \epsilon_{\alpha,\beta}\partial_{\alpha}b^{+}_{\beta})
+\frac{m}{2 \bar{\rho}}(\partial_{0}b^{+}_{\alpha})^{2}  + \frac
{1}{2}(\nabla \times \vec{b}^{+})V_{+}( \vec{q} ) (\nabla \times
\vec{b}^{+})
              \nonumber   \\
 & + & \frac{m}{2 \bar{\rho}}(\partial_{0}b^{-}_{\alpha})^{2} -\frac{1}{4
\pi}\frac{\bar{\rho}}{m}(\pi J^{v-}_{0})ln|x-y|(\pi
J^{v-}_{0})+\frac {1}{2}(\nabla \times \vec{b}^{-}) V_{-}( \vec{q} )
(\nabla \times \vec{b}^{-})
\end{eqnarray}

   The corresponding Hamiltonian is:
\begin{eqnarray}
{\cal H} &=& \frac{1}{2}\frac{\bar {\rho}}{m}q^{2}\theta_{+} ( -
\vec{q} )\theta_{+} (\vec{q}) +\frac{1}{2}\frac{\bar {\rho}}{m}(\pi
J^{v+}_{0}(-\vec{q})-4 \pi \Pi_{+}(-\vec{q}))\frac{1}{q^{2}}(\pi
J^{v+}_{0}(\vec{q}) -4 \pi \Pi_{+}(\vec{q}))+  2 \Pi_{+}(-\vec
{q})V_{+}( \vec{q} )  \Pi_{+}(\vec {q})
     \nonumber \\
& + &\frac{1}{2}\frac{\bar {\rho}}{m}q^{2}\theta_{-} ( - \vec{q}
)\theta_{-} (\vec{q})+\frac{1}{2}\frac{\bar {\rho}}{m}(\pi
J^{v-}_{0}(-\vec{q}))\frac{1}{q^{2}}(\pi
J^{v-}_{0}(\vec{q}))+\frac{1}{2}\Pi_{-}(-\vec {q}) V_{-}( \vec{q} )
\Pi_{-}(\vec {q})
\label{decouple}
\end{eqnarray}
where $ [ \theta_{\pm} ( \vec{q} ), \Pi_{\pm} ( \vec{q}^{\prime} ) ]
= i \hbar \delta( \vec{q} + \vec{q}^{\prime} ) $.

From the above Hamiltonian in the bosonic representation, we can see
the charge sector and spin sector remain " decoupled "
\cite{decouple}. Due to the absence of the C-S term in the spin
sector, the inserted vortex only shifts the total density variable
in the charge sector, but does not couple to the relative density in
the spin sector, so the Hamiltonian in the spin sector remains the
same as the ground state one Eqn.\ref{spinh}, the corresponding
wavefunction remains the same as the ground state one in the spin
sector Eqn.\ref{spinw}. All the effects of the inserted vortex are
encoded in the charge sector. Again, neglecting the Coulomb
interaction $ V_{+}(q) $ in the long wavelength limit, we find that
the wavefunction in the charge sector is:
\begin{equation}
\Psi^{b}_{cqh} = \exp[
\frac{1}{2}\sum_{q}(\frac{1}{2}J^{v+}_{0}(-\vec{q})-
 \delta \rho_{+}(-\vec{q})) (-\frac{2 \pi }{q^{2}} ) (\frac{1}{2}J^{v+}_{0}(\vec{q}) - \delta
 \rho_{+}(\vec{q}))]
\end{equation}
     Transforming to the coordinate space and setting $ J^{v+}_{0}(x)=J^{v-}_{0}(x)=-\delta(x)
     $ \cite{decouple} lead to:
\begin{eqnarray}
\Psi^{b}_{cqh} &=& \exp[ \frac{1}{2} \int dx dy ( \frac{1}{2}
\delta(\vec{x}) + (\sum_{i}\delta({\vec{x}}-z_{i}) +
 \delta({\vec{x}}-w_{i})-\bar{\rho}))ln|x-y|(\frac{1}{2} \delta(\vec{y}) +
 (\sum_{i}\delta({\vec{y}}-z_{i})+\delta({\vec{y}}-w_{i})-\bar{\rho}))]     \nonumber  \\
&=&\prod^{N_{1}}_{i}|z_{i}|^{\frac{1}{2}}\prod^{N_{2}}_{i}|w_{i}|^{\frac{1}{2}}\prod^{N_{1}}_{i=1}
\prod^{N_{2}}_{j=1} |z_{i}-w_{j} | \prod^{N_{1}}_{ i< j }
|z_{i}-z_{j}|  \prod^{N_{2}}_{ i< j } | w_{i}-w_{j}| \label{qhm}
\end{eqnarray}

   The SGT for the quasi-hole could be different from that for the
   groundstate. If one inserts a vortex at the origin at the layer
   1 in the {\em boson} Lagrangian Eqn.\ref{boson}, in order to recover the original
   electronic Hamiltonian Eqn.\ref{first}, $ U_{0} $ in Eqn.\ref{singg} is needed to remove the C-S term,
   an additional SGT $ U_{v1} $ is needed to remove the effects of
   the inserted vortex. In the first quantization, it is easy to
   show that \cite{slqht}:
\begin{equation}
   U_{v1}= e^{i \sum_{i} arg \frac{z_{i}}{ |z_{i} | } }= e^{ \sum_{i} ln \frac{z_{i}}{ |z_{i} | } }
          =\prod_{i} \frac{z_{i}}{ |z_{i} | }
\label{v1}
\end{equation}
   The total SGT for the quasi-hole at the layer 1 is $ U_{qh}= U_{0}
   U_{v1} $.

   Performing the SGT on Eqn.\ref{qhm}, we get the quasi-hole wavefunction,
\begin{equation}
  \Psi_{qh}(z,w)= ( \prod^{N_{1}}_{i}z_{i}) \prod_{i} |\frac{w_{i}}{z_{i}}|^{1/2}
   \Psi_{0}(z,w)
\label{qh}
\end{equation}
   where $ \Psi_{0}(z,w) $ is the ground state wavefunction Eqn.\ref{totalg} and
   $ N_{1}=N_{2} = N $ in the balanced case. Note that there is no
   singularity at the origin.

   Note that even in the $ d \rightarrow 0 $ limit, Eqn.\ref{qh}
   differs from both the " quasi-hole "
   and the " meron " wavefunction listed in Eqn.\ref{prime}. This
   should not cause any problem. It is known that $ (111) $ wavefunction
   is the exact wave function in the $ d \rightarrow o $ limit. But
   both both the " quasi-hole "
   and the " meron " wavefunction listed in Eqn.\ref{prime} make sense only at finite distances.
   It is known that the lowest energy excitation at $ d=0 $ is a
   skymion carrying charge 1, while the  " quasi-hole "
   and the " meron "  carry total charge $ 1/2 $, so they are not
   valid excitations anymore at $ d=0 $. We conclude the quasi-hole
   Eqn.\ref{qh} make sense only at finite $ d $. It is not
   interesting to take $ d \rightarrow 0 $ limit to this equation \cite{boring}.

   Compared to the " quasi-hole "
   and the " meron " wavefunction listed in Eqn.\ref{prime}, we can see that
   there are two modifications in the Eqn.\ref{qh}: (1) The ground
   state wavefunction is the correct one Eqn.\ref{totalg} instead of the $ (111) $ wavefunction.
   (2) The prefactor is different from both the "quasi-hole " and the " meron ". We
   expect this prefactor takes care of the strong interlayer
   correlations. All the wavefunctions in Eqn.\ref{prime} are built upon the $ (111)
   $ wavefunction. As
   suggested in \cite{wave} and explicitly shown in this paper, $
   (111) $ wavefunction is not even qualitatively correct at any
   finite $ d $. As shown in \cite{wave}, although
   prefactors are not important in SLQH due to the gap in the bulk, they maybe crucial in BLQH
   due to the gapless mode in the interlayer correlations.
   The two factors maybe responsible for the " quasi-hole
   "'s charge distribution spreading over the whole system and its
   energy diverges linearly with the area of the system.
   Although, the "meron " wavefunction's energy is only logarithmically divergent,
   it ignores the strong interlayer correlations, so it is not a
   good trial wavefunction either.
   We propose that the quasi-hole wave function Eqn.\ref{qh} not only has logarithmically
   divergent energy, well localized charge distribution, but also
   correct interlayer correlations.

\section{ A pair of quasihole excitations with charge 1 }

 Now we put two vortices into the BLQH system. One is on the top layer
 at $z_{0}$ and the other is on the bottom layer at $w_{0}$. The only
 change from the quasi-hole calculation in the last section is
 $ J^{v \pm}_{0}=\delta(\vec{x}-z_{0}) \pm \delta(\vec{x}-w_{0})$ \cite{decouple}.
 Again, due to the lack of the CS term in the spin sector,
 the Hamiltonian in the spin sector remains the same as the ground state
 one Eqn.\ref{spinh}, so the wavefunction  is not affected at all by
 the insertion of the two vortices and remains the same as the ground state
 one in the spin sector Eqn.\ref{spinw}. All the effects of the
 inserted two vortices are encoded in the charge sector.
 The wavefunction in the charge sector in the bosonic picture is:
\begin{eqnarray}
\Psi^{b}_{c1} &=& \exp( \frac{1}{2} \int dxdy [\frac{1}{2} (
 \delta(\vec{x}-z_{0})+\delta(\vec{x}-w_{0}))
 +(\sum_{i}\delta({\vec{x}}-z_{i})+\delta({\vec{x}}-w_{i})-\bar{\rho})]ln|x-y|
 \nonumber \\
  &&[(\frac{1}{2}(\delta(\vec{y}-z_{0})+\delta(\vec{y}-w_{0}))+(\sum_{i}\delta({\vec{y}}-z_{i})+
  \delta({\vec{y}}-w_{i})-\bar{\rho}))]
  \nonumber \\
&&=\prod^{N_{1}}_{i}|z_{i}-z_{0}|^{\frac{1}{2}}\prod^{N_{1}}_{i}|z_{i}-w_{0}|^{\frac{1}{2}}
\prod^{N_{2}}_{i}|w_{i}-z_{0}|^{\frac{1}{2}}\prod^{N_{2}}_{i}|w_{i}-w_{0}|^{\frac{1}{2}}
\label{pc}
\end{eqnarray}

  It is easy to see that in the bosonic picture, the above
  wavefunction in the charge sector is symmetric under $ z_{i} \leftrightarrow w_{i} $ or
  $ z_{0} \leftrightarrow w_{0} $ separately. This is under
  expectation, because the two layers are completely symmetric in
  the charge sector.

  Just like deriving $ U_{v1} $ for the quasi-hole, we can get $
  U_{v2} $ for a pair of vortices inserted at $ z_{0} $ in top layer and $ w_{0} $ at the bottom layer:
\begin{equation}
   U_{v2}= \prod_{i} ( \frac{z_{i}-z_{0}}{ |z_{i}-z_{0} | } ) ( \frac{w_{i}-w_{0}}{ |w_{i}-w_{0} |
   })
\label{v2}
\end{equation}
   The total SGT for the meron pair is $ U_{pair}= U_{0} U_{v2}$.

   Performing the SGT on Eqn.\ref{pc}, we get the wavefunction for a pair of
   quasihole:
\begin{equation}
  \Psi_{pair}(z,w;z_{0},w_{0})= ( \prod_{i} ( z_{i}- z_{0} ) ( w_{i}- w_{0} )
                     \prod_{i} |\frac{( z_{i} -w_{0} )( w_{i}-z_{0} ) }
                      { ( z_{i}-z_{0} ) ( w_{i}-z_{0} ) } |^{1/2}  \Psi_{0}(z,w)
\label{pair}
\end{equation}
   where $ \Psi_{0}(z,w) $ is the ground state wavefunction Eqn.\ref{totalg} and
   $ N_{1}=N_{2} = N $ in the balanced case. Note that the pair
   wavefunction is not symmetric under $ z_{i} \leftrightarrow w_{i} $ or
   $ z_{0} \leftrightarrow w_{0} $ separately anymore. This is
   because the SGT  $ U_{v2} $ Eqn.\ref{v2} is not. Of course, it is still
   symmetric under $ z_{i} \leftrightarrow w_{i}, z_{0} \leftrightarrow w_{0}
   $ simultaneously.

   If we insert the two vortices at the same point, namely, putting $
   z_{0}=w_{0}=0 $ in the above equation, as expected, we get:
\begin{equation}
  \Psi_{pair}(z,w,0) =( \prod_{i}  z_{i}  w_{i} ) \Psi_{0}(z,w)
\label{qh0}
\end{equation}

    This corresponds to insert a single vortex through the two
    layers. In contrast to the quasi-hole excitation Eqn.\ref{qh},
    Eqn.\ref{qh0} carries charge 1 and remains a valid wavefunction even at $ d=0
    $. Indeed, in the $ d \rightarrow 0 $ limit, it recovers the " meron pair " wavefunction listed in Eqn.\ref{prime}.
    If one splits the whole vortex, it will evolve into the
    pair wavefunction Eqn.\ref{pair}.

   The "pair meron" wavefunction built on $ (111) $ is listed in \ref{prime}( essentially Eqn.(110) in Ref.10 ).
   It was shown that its energy $ E_{pair} \sim | z_{0}-w_{0} |^{2} $ instead of
   logarithmically as naively expected,
   because the charges are extended between $ z_{0} $ and $ w_{0} $.
   Similar to the quasi-hole  wavefunction Eqn.\ref{qh},
   there are two modifications in the Eqn.\ref{pair}: (1) The ground
   state wavefunction is the correct one Eqn.\ref{totalg}  (2) The prefactor is
   different. We expect this prefactor takes care of the strong interlayer
   correlations between the two vortices, the pair wave function not only has logarithmically
   divergent energy, well localized charge distribution, but also
   correct interlayer correlations.

\section{ Instability in the wavefunction as the distance increases }

    When the distance is sufficiently small, the BLQH is in the ESF
    phase, we expect the ground state, quasi-hole and pair
    wavefunctions Eqns.\ref{totalg},\ref{qh},\ref{pair} only hold in the
    ESF phase. When the distance becomes sufficiently large, the two
    layers become two weakly coupled Fermi liquid ( FL ) layers.
    All these wavefunctions completely break down. New set of
    wavefunctions are needed.
    Although the ESF phase and FL phase at the two extreme distances are well established,
    the picture of how the ESF phase evolves into the two weakly-coupled FL states was not clear, namely,
    the nature of the intermediate phase at $  d_{c1} < d < d_{c2} $ was still under debate.
    Recently,  starting from the well studied excitonic superfluid (ESF) state, as distance increases,
    one of the authors found  \cite{cbprl} that
   the instability driven by magneto-roton minimum collapsing at a finite wavevector in the pseudo-spin channel
   leads to the formation of a pseudo-spin density wave (PSDW) at
   some intermediate distances. He constructed a quantum
   Ginsburg-Landau theory  to study the transition from the ESF to the
   PSDW and analyze in detail the properties of the PSDW.
   He showed that a square lattice is the favorite lattice.

   As shown in \cite{cbprl},  it is the original
   instability in $ V_{-}( q )= a- bq+cq^{2} $ which leads to the magneto-roton
   minimum in the Fig.1a. By looking at the two conditions $ V_{-}(\vec{q})|_{q=q_{0}}=0 $
   and $ \frac{ d V_{-}(\vec{q})}{ d q }|_{q=q_{0}}=0 $,
   it is easy to see that $ V_{-}(q) $ indeed has the shape shown in
   Fig.1b. When $ b \sim d^{2} < b_{c} = 2 \sqrt{ac} \sim d $,
   the minimum of $ V_{-}(q) $ at $ q= q_{0}=\sqrt{a/c} \sim d $ has a gap,
   the system  is in the ESF state, this is always the case when the distance $ d $ is
   sufficiently small. However, when $ b = b_{c} $, the minimum collapses  and  $ S(q) $ diverges at $ q= q_{0}
   $, which signifies the instability of the ESF to the pseudo-spin density wave (PSDW) formation.
   When $ b \sim d^{2} > b_{c} = 2 \sqrt{ac} \sim d $, the minimum
   drops to negative, the system gets to the PSDW state, this is always the case when the distance $ d $ is
   sufficiently large.

\vspace{0.25cm}

\epsfig{file=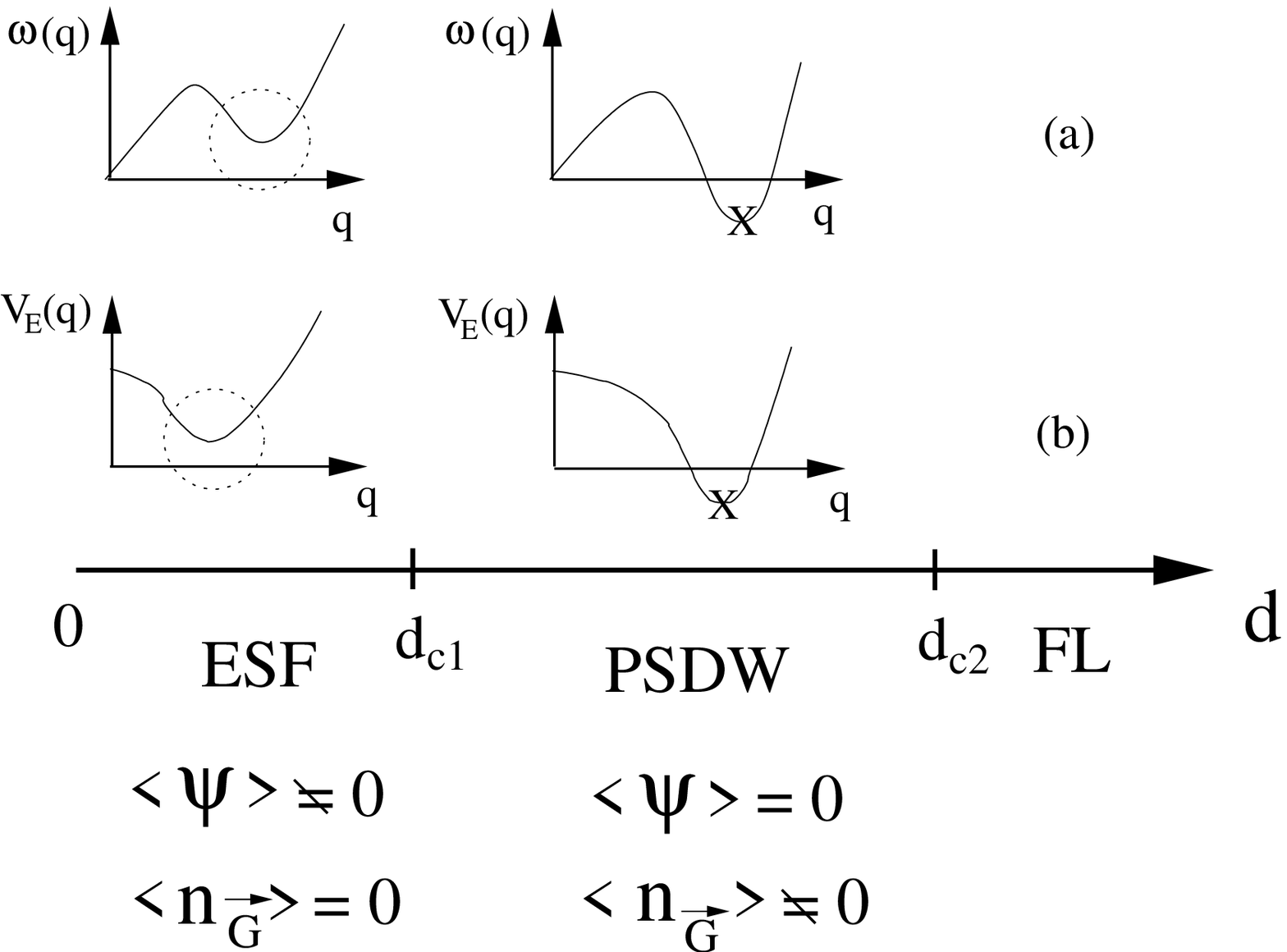,width=3.1in,height=2.0in,angle=0}

\vspace{0.25cm}

{\footnotesize {\bf Fig.1:}  The zero temperature phase diagram in
the balanced case as the distance between the two layers increases.
ESF where $ < \psi> \neq 0, < n_{\vec{G} } >=0 $ stands for
excitonic superfluid, PSDW where $ < \psi> = 0, < n_{\vec{G} } >
\neq 0 $ stands for pseudo-spin density wave phase, FL stands for
Fermi Liquid. (a) Energy dispersion relation $ \omega(q) $ in these
phases. (b) $ V_{E}(q) $ in these phases. The cross in the PSDW
means the negative minimum value of $ V_{E}( q ) $ is replaced by
the PSDW. The two order parameters were defined in \cite{cbtwo}. In
reality, the instability happens {\sl before } the minimum
collapses. }

\vspace{0.25cm}

   We can easily see the instability from the ground state
   wavefunction Eqn.\ref{totalg}. The only distance dependence in the
   charge sector is encoded in the $ V_{+}(q) $ in the bosonic Hamiltonian
   in the charge sector Eqn.\ref{ch}, but this dependence is ignored
   in the $ (111) $ wavefunction in the charge sector. The $ d $ dependence in $ V_{+} $
   is smooth anyway. Essentially
   all the distance dependence is encoded in the ground state wavefunction in the
   spin sector Eqn.\ref{spinw}. As can be seen from Fig.1, the
   instability happens at $ q=q_{0} $ where $ V_{-} $ becomes
   negative, but the spin stiffness $ \rho_{E} $ remains {\em
   non-critical } through the ESF to PSDW transition.
   As $ d \rightarrow d^{-}_{c1}
   $, the sum over $ q $ in Eqn.\ref{spinw} becomes dominated by the regime $ q \sim
   q_{0} $. When $ d_{c1} < d < d_{c2} $, Eqn.\ref{spinw} breaks
   down. A new wavefunction to describe the translational symmetry
   breaking PSDW state is needed. Some trial wavefunctions are
   proposed in \cite{trial}. It would be interesting to derive the
   new wavefunction of the PSDW state from the CB theory.

\section{Conclusions }

  BLQH differs from the SLQH by its symmetry breaking ground state
  and associated neutral gapless mode in the pseudo-spin sector.
  Due to the gapless mode in the bulk, the groundstate wavefunctions could be
  considerably different from the well known $ (111) $ wave
  function \cite{moon,wave}. The low energy excited states could
  also be sensitive to details such as normalization factors.
  One important problem is to find good trial wavefunctions for the ground state
  and low energy excited states. We investigated this important open
  problem from the CB theory developed previously in \cite{cbprl,cbtwo} to study BLQH systematically.
  We derived the ground state, quasi-hole and a pair of quasi-hole
  wavefunctions from CB theory and its dual action by the following procedures:
  We first performed the singular gauge transformation Eqn.\ref{singb} to transform a fermionic problem into
  a bosonic problem, then found that
  all the wavefunctions  in the bosonic picture are always the product of two parts, one part in the charge sector and
  the other in the spin sector. All the distance dependence are encoded in the
  spin part, while all the excitations only happen in the charge sector.
  After transforming back to the original electron picture by proper
  inverse SGT's, we get the final wavefunctions in the electron
  coordinates. We found that the inverse SGT's are different in
  ground state, meron and a pair of merons. By considering the
  differences carefully, we derived all these wavefunctions in the original electronic picture a systematical way.

  At any finite $ d $, the ground state wavefunction in
  the charge sector is the same as the $ (111) $ wavefunction, while
  that in the spin sector is highly nontrivial due to the gapless mode.
  So the total groundstate wavefunction differs from the
  well known $ (111) $ wavefunction at any finite $ d $.
  In the bosonic picture, when inserting vortices in the ground state, the spin
  part remains the same due to the lack of CS term in this sector, while the charge part changes accordingly.
  However, due to the insertion of vortices, in order to recover the original electronic problem,
  the inverse SGT differs from that in the ground state.
  After transforming back to the original electron problem by
  the inverse  SGT's, we showed that the quasi-hole and a pair of quasi-holes wavefunctions
  contain non-trivial normalization factors as shown in
  \ref{qh},\ref{pair}.  We expect  that the quasi-hole and pair wave function not only has logarithmically
  divergent energy, well localized charge distribution, but also
  correct interlayer correlations.  It is important to test these
  trial wavefunctions by QMC simulations performed in \cite{wave}
  for the states listed in Eqn. \ref{prime}. We also investigated the
  instability encoded in the spin sector which leads to the PSDW
  solid formation proposed in \cite{cbprl,cbtwo}. Because the CB
  field theory has been used to describe the tri-layer quantum Hall
  systems very successfully, the analysis in this paper can be easily extended to derive the
  wavefunctions in the TLQH  \cite{tricb}.

   It is well known that CB approach is  not a Lowest Landau Level (
    LLL ) approach \cite{moon,cbtwo}, it is very difficult to
   incorporate the LLL projection into the CB approach. This may be
   partially responsible for the spin part of the ground state wavefunction
   Eqn. \ref{spinw1} not in the LLL level. But as explained below Eqn.\ref{totalg},
   this should not be too worrisome, because
   Eqn.\ref{spinw1} works only in long distance anyway. As shown in \cite{wave} and listed
   in Eqn.\ref{prime}, in the long wavelength limit, the meron wavefunction's normalization factor
   contains modulus which is not in the LLL either.
   How to getting precise short distance behaviors of these
   wavefunctions from the CB theory remains an
   open problem.

   J. Ye thanks S. C. Zhang for helpful discussions. He also thanks
   Yu Lu and Xiang Tao for the hospitality during his visit to the
   Institute for Theoretical Physics of China.


\begin{references}
\bibitem{compare} As shown in detail in \cite{cbtwo}, the
functional form achieved from the Composite Boson theory is the same
at that achieved from the microscopic LLL+HF approach. But the
coefficients should be taken from the LLL+HF approach.
\bibitem{slqht} The same transformation is needed to transform the
modulus of the quasi-hole wavefunction in SLQH at Laughlin's series
$ \nu=1/m $ with $ m$ odd  $ |\Psi_{qh}|= \prod^{N_{1}}_{i} | z_{i}
| |\Psi_{m} | $ back to its final form $ \Psi_{qh}=
\prod^{N_{1}}_{i} z_{i} \Psi_{m} $. To the best of the authors'
knowledge, this very subtle and important point was not explicitly
spelled out in any previous work.
\bibitem{boring} It was shown in \cite{wave} that although the " quasi-hole " and the " meron " wavefunction listed in
  Eqn.\ref{prime} are completely different at any finite distance,
  they have finite comparable expectation energies at $
  d=0 $,  So we expect
  the quasi-hole's  expectation energy in Eqn.\ref{qh} is comparable to
  both at $ d=0 $. However, as explained in the main text, all
  these wavefunctions are not interesting anymore at $ d=0 $.
\end{references}
\end{document}